\documentclass[english,journal]{extarticle}
\usepackage{color}
\usepackage{subfigure}
\usepackage{graphicx}
\usepackage{float}
\usepackage{amsthm}
\usepackage{amsmath}
\usepackage{amssymb}
\usepackage{anysize}
\usepackage{parskip}
\usepackage{rotating}
\usepackage[unicode=true,pdfusetitle,
 bookmarks=true,bookmarksnumbered=false,bookmarksopen=false,
 breaklinks=false,pdfborder={0 0 0},backref=false,colorlinks=true]
 {hyperref}
\marginsize{0.8in}{0.8in}{0.7in}{0.7in}

\makeatletter

\usepackage{xy}
\xyoption{matrix}
\xyoption{frame}
\xyoption{arrow}
\xyoption{arc}

\usepackage{ifpdf}
\ifpdf
\else
\PackageWarningNoLine{Qcircuit}{Qcircuit is loading in Postscript mode.  The Xy-pic options ps and dvips will be loaded.  If you wish to use other Postscript drivers for Xy-pic, you must modify the code in Qcircuit.tex}
\xyoption{ps}
\xyoption{dvips}
\fi

\entrymodifiers={!C\entrybox}

\newcommand{\ket}[1]{{\left\vert{#1}\right\rangle}}
\newcommand{\qw}[1][-1]{\ar @{-} [0,#1]}
\newcommand{\qwx}[1][-1]{\ar @{-} [#1,0]}

\newcommand{\gate}[1]{*+<.6em>{#1} \POS ="i","i"+UR;"i"+UL **\dir{-};"i"+DL **\dir{-};"i"+DR **\dir{-};"i"+UR **\dir{-},"i" \qw}

\newcommand{\control}{*!<0em,.025em>-=-<.2em>{\bullet}}

\newcommand{\ctrl}[1]{\control \qwx[#1] \qw}

\newcommand{\targ}{*+<.02em,.02em>{\xy ="i","i"-<.39em,0em>;"i"+<.39em,0em> **\dir{-}, "i"-<0em,.39em>;"i"+<0em,.39em> **\dir{-},"i"*\xycircle<.4em>{} \endxy} \qw}

\newcommand{\multigate}[2]{*+<1em,.9em>{\hphantom{#2}} \POS [0,0]="i",[0,0].[#1,0]="e",!C *{#2},"e"+UR;"e"+UL **\dir{-};"e"+DL **\dir{-};"e"+DR **\dir{-};"e"+UR **\dir{-},"i" \qw}
\newcommand{\ghost}[1]{*+<1em,.9em>{\hphantom{#1}} \qw}

\newcommand{\rstick}[1]{*!L!<-.5em,0em>=<0em>{#1}}
\newcommand{\lstick}[1]{*!R!<.5em,0em>=<0em>{#1}}

\newcommand{\Qcircuit}{\xymatrix @*=<0em>}

\theoremstyle{plain}

\theoremstyle{definition}
\newtheorem{defin}{Definition}

\theoremstyle{remark}

\newtheorem{lemm}{Lemma}

\newcommand {\apgt} {\ {\raise-.5ex\hbox{$\buildrel>\over\sim$}}\ }
\newcommand {\aplt} {\ {\raise-.5ex\hbox{$\buildrel<\over\sim$}}\ }

\begin{document}

\title{Use of global interactions in efficient quantum circuit constructions}

\author{\small{Dmitri Maslov$^{1}$, Yunseong Nam$^{2,3}$} \\
{\small\it $^1$ National Science Foundation, Arlington, VA 22230, USA} \\
{\small\it $^2$ QuICS, University of Maryland, College Park, MD 20742, USA} \\
{\small\it $^3$ IonQ Inc., College Park, MD 20740, USA} \\
{\small\tt \href{mailto:dmitri.maslov@gmail.com}{dmitri.maslov@gmail.com}, \href{mailto:nam@ionq.co}{nam@ionq.co}}\\
}

\maketitle

\begin{abstract} 
In this paper we study the ways to use a global entangling operator to efficiently implement circuitry common to a selection of important quantum algorithms.  In particular, we focus on the circuits composed with global Ising entangling gates and arbitrary addressable single-qubit gates.  We show that under various circumstances the use of global operations can substantially improve the entangling gate count. 
\end{abstract}


\section{Introduction}
\label{SecI}

Trapped atomic ions \cite{DLF16} and superconducting circuits \cite{www:IBM} are two examples of quantum information processing (QIP) approaches that have delivered small yet already universal and fully programmable machines.  In superconducting circuits qubit interactions are enabled through custom designed electronic hardware involving Josephson junctions and microwave resonators \cite{www:IBM}.  Different interactions can be controlled individually to invoke the two-qubit gates.  A global coupling, however, would not necessarily be natural to such a system, due to the difficulty of placing and connecting $O(n^2)$ individual resonators in the same area as $n$ qubits. This said, it is possible to couple Josephson junction qubits to a single resonator mode, thereby enabling global interactions \cite{DS13}.  In trapped ion QIP, on the other hand, global interactions are more naturally realized as an extension of common two-qubit gate interactions \cite{SM99, SM00, MSJ00, MMNS16, MKH09}.  In fact, the ability to implement arbitrary selectable two-qubit interactions generally requires a higher level of control, with individually focused external fields addressing each qubit \cite{DLF16}.  Given the ease of implementing a global interaction over these two leading QIP approaches, we consider the use of global entangling gates, particularly applied to the trapped ions technology.  We note that the results are technology-independent and therefore apply to any QIP approach, so long as proper global entangling operations are constructible.

One particular interaction available in the trapped ions approaches \cite{DLF16, MMNS16, MKH09} to quantum computing is the so-called Molmer-Sorensen gate \cite{MS99}, also known as the XX coupling or Ising gate. To achieve computational universality, Molmer-Sorensen gate (either local addressable or global) is complemented by arbitrary single-qubit operations.  These may come in different flavors, including the addressable $R(\theta,\phi)$ rotations \cite{DLF16} of which at most two are needed to implement arbitrary single-qubit gate \cite{M17}, or the addressable RZ rotation, which together with global RX and RY rotations also gives the single-qubit universality \cite{MMNS16, MKH09}.  Depending on the specifics, the control apparatus may allow the application of an XX gate to a selectable pair of qubits \cite{DLF16}, globally \cite{SM99, SM00, MSJ00, MMNS16, MKH09}, or globally to a subset of qubits \cite{Monroe-PC,Blatt14}.  We furthermore note that the existing control apparatus described in reference \cite{DLF16} allows the application of the global Molmer-Sorensen (GMS) gates \cite{Monroe-PC}, however, to date, this approach has not been studied in detail.  In each case above, XX gate comes at a higher cost (expressed in terms of the duration and/or average fidelity) compared to the single-qubit gates.

In this paper, we focus on minimizing the number of times an XX gate is called---be it addressable local or global, thereby targeting the most expensive resource in quantum computations using trapped ions QIP.  Specifically, we center our efforts on finding the instances of quantum computations that admit a more efficient implementation using global entangling gates compared to what may be accomplished using local entangling gates.  Given that the control by global entangling operators applies a certain operation to multiple data, it can be thought of as a quantum analogue of classical SIMD (Single Instruction, Multiple Data) architecture.  Our goal in this paper is thus to demonstrate practical advantages of quantum SIMD architecture beyond those examples already known.

Previous work demonstrated how to implement the parity function (fan-in gate in our terminology) using a constant number of two global entangling pulses \cite{ZZY05}.  We revisit the implementation of fan-in in Subsection \ref{SecIIIa}, since it is relevant to our more advanced constructions.  Reference \cite[Figure~5]{BNO17} shows a two-GMS gate construction of the number excitation operator used in quantum chemistry simulations \cite{Qchem2,Qchem}.  References \cite{MMNS16, Ivanov} study the ways to implement quantum algorithms efficiently on a trapped ion quantum computer with the two-qubit gates enabled by the global entangling operator, concentrating on the case featuring anywhere between two to four qubits.  Reference \cite{M17} focuses on quantum circuit compiling in the scenario when local addressable two-qubit gates are available.  Reference \cite{LW17} revisits the two-GMS gate parity measurement implementation of \cite{ZZY05} and reduces the number of global pulses needed to just one (this construction can be inferred from Fig.~\ref{fig2}), and shows how to measure the eigenvalue of a product of Pauli matrices using only a constant number of global entangling pulses. In contrast, here, we determine a set of important quantum circuits, focusing on the computations of arbitrary size, that can be accomplished using fewer entangling pulses in cases when global entangling control is available.  The new circuits developed in our work include stabilizer circuits, Toffoli-4 gate, Toffoli-$n$ gate, Quantum Fourier Transformation, and Quantum Fourier Adder circuits, thereby substantively extending the set of known efficient circuitry based on the global entangling pulse.  The results are directly accessible for implementation over trapped ions approaches featuring global control, and make a case for mixed local/global entangling control.

Computational universality of the control given by selectable two-qubit couplings and arbitrary single-qubit gates was the subject of an early foundational study establishing the upper bound of $O(n^34^n)$ and the lower bound of $\Omega(4^n)$ on the number of the CNOT gates required to implement an arbitrary unitary \cite{BBCD96}.  The upper bound was later improved to $O(4^n)$ in \cite{MVBS04, SBM06}, at which point it asymptotically met the lower bound, settling the question of asymptotically optimal control by the entangling CNOTs gates.  For logical-level fault tolerant circuits one more step is needed---specifically, that of decomposing all gates into a discrete fault-tolerant library, such as the one given by the Clifford and T gates.  With the CNOT being a Clifford gate, the remaining step on top of asymptotically optimal constructions of \cite{MVBS04, SBM06} is to decompose arbitrary single-qubit unitaries into Clifford+T circuits.  Euler angle decomposition may be used to express arbitrary single-qubit unitary as a circuit with no more than three axial rotations \cite{NC}, and $z$-rotations can be synthesized optimally as single-qubit Clifford+T circuits \cite{KMM13,RS16}.  Given additional resources such as in-circuit measurement and classical feedback, even better solutions exist \cite{BRS15}.  

The upper bound of $O(4^n)$ on the number of CNOT gates \cite{MVBS04, SBM06} gives rise to the upper bound of $O(4^n)$ on the number of GMS gates, since it can be easily established that a CNOT gate can be obtained using no more than constantly many GMS gates.  Indeed, a 4-GMS implementation of the CNOT gate can be obtained by applying the two-GMS construction illustrated in Fig.~\ref{fig1} to $n$ qubits with $\chi=\pi/2$, and then again to $n{-}1$ qubits with $\chi=-\pi/2$, selecting one specific qubit-to-qubit interaction that remains active.  With the use of the maximal size GMS gates, this may be a slightly larger construction, relying on Fig.~\ref{fig12} to express smaller GMS gates in terms of the maximal size GMS gate, but one with constantly many GMS gates nonetheless.

In most practical cases, one may desire to implement a specific well-structured computation, and those frequently come with known implementations relying on fewer than $O(4^n)$ entangling gates.

Control by local addressable operations is clearly easier to work with as far as implementing quantum computations is concerned, since most quantum algorithms are expressed in terms of local operations. Secondly, the number of arbitrarily selectable two-qubit operations, $\frac{(n-1)n}{2}$, for an $n$-qubit computation (recall that the XX coupling does not distinguish between gate's control and its target), is higher than $1$, being the number of individual full-size global gates.  Thirdly, an arbitrary circuit over two-qubit local control experiences only a constant factor blow up if needs be implemented as a circuit over global control (this is no more true if global control needs be expressed in terms of local control).  These observations suggest that the local control is overall more nimble when it comes to implementing arbitrary quantum algorithms.  However, it is not always the case that the implementations using local addressable gates are more efficient compared to those over global entangling operators.  Indeed, it is known how to implement the $3$-qubit Toffoli gate with only three size-$3$ GMS gates \cite{MMNS16, MKH09}, whereas the best known implementation over two-qubit local addressable control requires five entangling gates \cite{NC}.  Motivated by this example, we look into what other important unitary transformations benefit from the control by global gates.

\section{Global MS Gate}\label{SecII}

A local MS gate (XX) \cite{MS99}, acting on $i^\text{th}$ and $j^\text{th}$ qubits, is defined as
\begin{align}
\label{MS}
\text{XX}_{ij}(\chi_{ij}) 
&= e^{-i(\hat{\sigma}_{x}^{(i)}+\hat{\sigma}_{x}^{(j)})^2\chi_{ij}/4} 
= e^{-i\hat{\sigma}_{x}^{(i)}\hat{\sigma}_{x}^{(j)}\chi_{ij}/2} \nonumber \\
&= \left(\begin{matrix}
&\cos(\chi_{ij}/2) \quad &0 \quad &0 \quad &-i\sin(\chi_{ij}/2) \\
&0 \quad &\cos(\chi_{ij}/2) \quad &-i\sin(\chi_{ij}/2) \quad &0 \\
&0 \quad &-i\sin(\chi_{ij}/2) \quad &\cos(\chi_{ij}/2) \quad &0 \\
&-i\sin(\chi_{ij}/2) \quad &0 \quad &0 \quad &\cos(\chi_{ij}/2)
\end{matrix}\right),
\end{align}
where $\hat{\sigma}_{x}^{(i)}$ denotes the Pauli-$x$ operator
acting on $i^\text{th}$ qubit. In comparison, a global MS (GMS) gate
for an $n$-qubit system is defined according to the equation 
\begin{align}
\label{GMS}
{\rm GMS}(\chi_{12},\chi_{13}, \ldots, \chi_{1\;n},\chi_{23}, \ldots, \chi_{n-1\;n}) 
&= \exp\left( -i\sum_{i=1}^{n}\sum_{j=i+1}^{n} (\hat{\sigma}_{x}^{(i)} + \hat{\sigma}_{x}^{(j)})^2\chi_{ij}/4\right) \nonumber \\
&= \exp\left( -i\sum_{i=1}^{n}\sum_{j=i+1}^{n} \hat{\sigma}_{x}^{(i)} \hat{\sigma}_{x}^{(j)}\chi_{ij}/2\right),
\end{align}
which is equivalent to the application of local XX gates to all $\frac{n(n-1)}{2}$ pairs of qubits for an $n$-qubit system. Since any two local XX gates always commute, the GMS gate is uniquely defined.  For simplicity, we will first focus on the GMS gate where $\chi_{12}=\chi_{13}=\ldots=\chi_{1n}=\chi_{23}=\ldots=\chi_{n{-}1\;n}$ \cite{SM99, SM00, MSJ00, MKH09}, and next consider other variants.

\begin{figure}
\begin{eqnarray*}
\Qcircuit @C=.5em @R=.8em @!R {
\lstick{1} & \qw	& \multigate{3}{{\rm GMS}4(\chi)} & \qw & \multigate{2}{{\rm GMS}3(-\chi)} & \qw\\
\lstick{2} &	 \qw	& \ghost{{\rm GMS}4(\chi)}           & \qw & \ghost{{\rm GMS}3(-\chi)} & \qw \\
\lstick{3} &	 \qw	& \ghost{{\rm GMS}4(\chi)}           & \qw & \ghost{{\rm GMS}3(-\chi)} & \qw \\
\lstick{4} &	 \qw	& \ghost{{\rm GMS}4(\chi)}           & \qw & \qw & \qw \\
}
&
\raisebox{-2.6em}{\hspace{1mm}=\hspace{1mm}}
&
\Qcircuit @C=.5em @R=.8em @!R {
& \qw & \qw & \qw & \multigate{3}{{\rm XX}(\chi)} & \qw & \rstick{1} \\
& \qw & \qw & \multigate{2}{{\rm XX}(\chi)} & \ghost{{\rm XX}(\chi)} & \qw & \rstick{2}\\
& \qw & \multigate{1}{{\rm XX}(\chi)} & \ghost{{\rm XX}(\chi)} & \ghost{{\rm XX}(\chi)} & \qw & \rstick{3}\\
& \qw & \ghost{{\rm XX}(\chi)} &\ghost{{\rm XX}(\chi)} & \ghost{{\rm XX}(\chi)} & \qw & \rstick{4} \\
}
\end{eqnarray*}
\caption{ \label{fig1}
Example of the usefulness of global gates.  GMS4 denotes a GMS gate defined according to (\ref{GMS}), applied to all four qubits shown in the figure. GMS3 denotes a three-qubit GMS gate, applied to qubit numbers 1, 2, and 3.  The common argument $\chi$ of the GMS gates specifies that all $\chi_{ij}$'s are equal to $\chi$.  The XX$_{ij}$($\chi$) gate denotes a local XX gate, applied to qubits $i$ and $j$ with the angle $\chi$, see (\ref{MS}).
}
\end{figure}
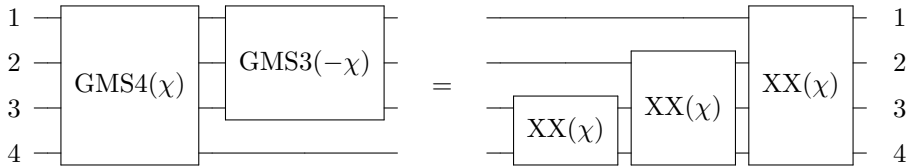

Intuitively, the availability of the GMS gate allows for an efficient implementation of a single-qubit--to--many-qubits coupling gate.  Consider, for instance, a 4-qubit system as shown in Fig.~\ref{fig1}.  Applying the GMS gate on all four qubits and then applying the GMS gate to the top three qubits with the negative sign of the rotation parameter, results in a selective set of the XX gates acting between qubit number 4 and the rest, as shown in Fig.~\ref{fig1} on the right.  This means that, together with the ability of leaving out a qubit of choice, we need only two (global) entangling operators to perform the desired transformation.  Note that because qubit number 4 participates in all three XX gates as shown in Fig.~\ref{fig1} on the right hand side, even with the possibility of parallel operations acting on disjoint pairs of qubits at least three time steps would be required if we restrict ourselves to the local XX couplings.

In the rest of the paper, we rely on the standard \cite{NC} single-qubit gates, including Hadamard ($H$ in formulas and circuit diagrams), axial rotations RX, RY, and RZ ($X$, $Y$, and $Z$ in circuit diagrams), as well as the two-qubit CNOT gate. 

\section{Efficient circuits using the GMS gate}
\label{SecIII}

In this section, we present a suite of quantum transformations, where GMS gates may be handily used to increase circuit efficiency. We lay out the specific implementation details by explicitly constructing corresponding quantum circuits, and compare them to those obtained using only local entangling gates to highlight the efficiency gain.

\subsection{Consecutive CNOTs: Single-Control Many-Target CNOT (fan-out), and Many-Control Single-Target CNOT (fan-in)}
\label{SecIIIa}

\begin{figure}
\begin{eqnarray*}
\Qcircuit @C=.7em @R=1.4em @!R {
& \ctrl{3} & \qw \\
& \targ    & \qw \\
& \targ    & \qw \\
& \targ    & \qw 
}
&
\raisebox{-3.3em}{\hspace{1mm}=\hspace{1mm}}
&
\Qcircuit @C=1em @R=1.4em @!R {
& \ctrl{1} & \ctrl{2} & \ctrl{3}  & \qw \\
& \targ    & \qw     & \qw       & \qw \\
& \qw     & \targ    & \qw       & \qw \\
& \qw     & \qw     & \targ      & \qw 
}
\raisebox{-3.3em}{\hspace{1mm}=\hspace{1mm}}
\Qcircuit @C=.4em @R=.4em @!R {
& \gate{Y\frac{\pi}{2}} 	& \multigate{3}{{\rm GMS}4(\pi/2)} 	& \qw & \qw  					       		&\gate{X{-}\frac{3\pi}{2}}	&\gate{Y{-}\frac{\pi}{2}} & \qw \\
& \qw               		& \ghost{{\rm GMS}4(\pi/2)}         & \qw & \multigate{2}{{\rm GMS}3(-\pi/2)}	&\gate{X{-}\frac{\pi}{2}} 	&\qw & \qw \\
& \qw                    	& \ghost{{\rm GMS}4(\pi/2)}         & \qw & \ghost{{\rm GMS}3(-\pi/2)}          &\gate{X{-}\frac{\pi}{2}} 	&\qw & \qw \\
& \qw                    	& \ghost{{\rm GMS}4(\pi/2)}         & \qw & \ghost{{\rm GMS}3(-\pi/2)}          &\gate{X{-}\frac{\pi}{2}} 	&\qw & \qw 
}
\end{eqnarray*}
\caption{\label{fig2}
Four-qubit case of multiple CNOT gates sharing a single control qubit and targeting the rest of the qubits. Only two GMS gates are required to implement a total of $n{-}1$ local XX gates, corresponding to $n{-}1$ CNOT gates.
}
\end{figure}
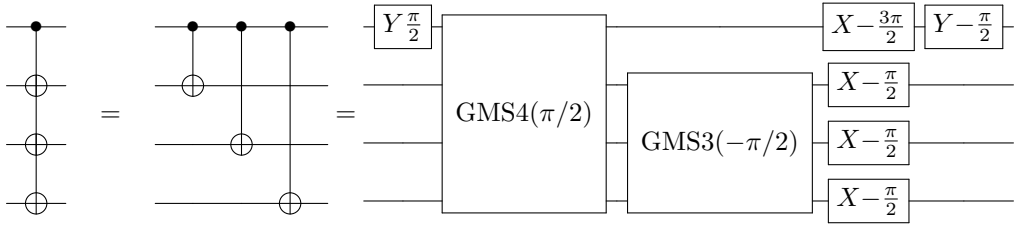

Consider a set of CNOT gates with a shared control qubit, also known as the fan-out gate.  As illustrated in Fig.~\ref{fig2} for the sample case of $n=4$, we can use a pair of GMS gates, together with single-qubit rotations $\text{RX}(\theta) = e^{-i\hat{\sigma}_x\theta/2}$ and $\text{RY}(\theta) = e^{-i\hat{\sigma}_y\theta/2}$, to implement the entire set of such $n{-}1$ CNOT gates. In particular, we require a total of two GMS gates, one over $n$ qubits with uniform angles $\pi/2$ and the other over $n{-}1$ qubits with the angle $-\pi/2$, singling out the control qubit.

An $n$-qubit fan-in gate (a set of CNOTs sharing a target) can be implemented as a layer of $n$ Hadamard gates, followed by the fan-out gate, followed by the second layer of $n$ Hadamard gates.  This means that an arbitrary size fan-in gate can too be implemented using a constant number of two GMS gates.  We note that these implementations were known to \cite{ZZY05, LW17} (fan-in was explicitly studied, and fan-out can be easily obtained from the fan-in).  Observe that to measure the outcome of the parity function on the top qubit (see Fig.~\ref{fig2}), the second GMS gate in the construction outlined in Fig.~\ref{fig2} needs not be applied, as it does not affect the qubit being measured.

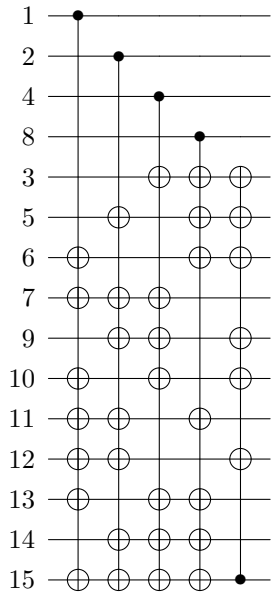
\begin{figure}[t]
\begin{eqnarray*}
\Qcircuit @C=.7em @R=.7em @!R {
\lstick{1}	&\ctrl{14}	&\qw      	&\qw		&\qw		&\qw		&\qw \\
\lstick{2}	&\qw     	&\ctrl{13}	&\qw		&\qw		&\qw		&\qw \\ 
\lstick{4}	&\qw     	&\qw      	&\ctrl{12}	&\qw		&\qw		&\qw \\
\lstick{8}	&\qw     	&\qw		&\qw		&\ctrl{11}	&\qw		&\qw \\
\lstick{3}	&\qw	   	&\qw     	&\targ		&\targ		&\targ		&\qw \\
\lstick{5}	&\qw     	&\targ   	&\qw		&\targ		&\targ		&\qw \\
\lstick{6}	&\targ   	&\qw	    &\qw	    &\targ		&\targ		&\qw \\
\lstick{7}	&\targ		&\targ		&\targ		&\qw    	&\qw		&\qw \\
\lstick{9}	&\qw	  	&\targ		&\targ		&\qw		&\targ		&\qw \\
\lstick{10}	&\targ   	&\qw       	&\targ  	&\qw   		&\targ		&\qw \\
\lstick{11}	&\targ   	&\targ		&\qw		&\targ  	&\qw 		&\qw \\
\lstick{12}	&\targ   	&\targ     	&\qw		&\qw    	&\targ 		&\qw \\
\lstick{13}	&\targ   	&\qw	    &\targ  	&\targ  	&\qw 		&\qw \\
\lstick{14}	&\qw   		&\targ    	&\targ  	&\targ  	&\qw 		&\qw \\
\lstick{15}	&\targ   	&\targ  	&\targ  	&\targ  	&\ctrl{-10} &\qw 
}
\end{eqnarray*}
\caption{\label{fig3}
Encoding circuit {\em Tdistill} of the $[[15,1,3]]$ code \cite{RHG07}, used to distil the $\ket{A}$ state.  It relies on a set of $34$ CNOT gates, that can be implemented using only $10$ GMS gates. 
}
\end{figure}

An immediate application of this efficient implementation ($n{-}1$ local XX gates replaced by a pair of GMS gates) may be observed, for instance, in stabilizer circuit constructions. 

Fig.~\ref{fig3} shows the encoding circuit for the 15-qubit Reed-Muller code \cite[Figure~12]{RHG07}.  This circuit containing a total of $34$ CNOT gates may be implemented with $5$ pairs of GMS gates, which would otherwise require $34$ local XX gates.  Since the $[[15,1,3]]$ encoding circuit is used to distil the $\ket{A}$ state \cite{RHG07}, its efficient GMS-enabled implementation may potentially be used to synthesize the logical-level $T$ gate efficiently, constituting an important optimization for fault-tolerant quantum computing.  We note, however, that GMS gates may be difficult to use fault-tolerantly \cite{BNO17}.

GMS gates can furthermore be employed to obtain an implementation of arbitrary $n$-qubit stabilizer unitary using at most $12n{-}18$ entangling pulses. To establish this, consider the 9-stage layered decomposition -C-P-C-P-H-P-C-P-C- of \cite{MR17}. Observe that two of the -C- stages (each corresponds to the CNOT-based circuits) are given by the upper triangular Boolean matrices. This means that each can naturally be implemented as a set of $n{-}1$ fan-out gates.  Of these, the smallest fan-out is the CNOT, and thus it can be implemented using a single GMS.  This means that the total number of GMS gates required to implement an upper/lower triangular linear reversible transformation is $2n{-}3$. The other two -C- stages are arbitrary linear transformations.  Using LU decomposition, each can be implemented as a circuit over $2(2n-3) = 4n-6$ GMS gates.  The total GMS count required to implement an arbitrary stabilizer unitary is thus $2(2n-3) + 2(4n-6) = 12n-18$. 

The number of GMS gates required to implement an arbitrary stabilizer unitary, $12n{-}18$, is significantly less than $\Omega(\frac{n^2}{\log n})$ of the two-qubit CNOT gates required to accomplish the same \cite{PMH08}.  The comparison, however, is not fair. This is because the number of different functions computed by the CNOT gates spanning $n$ qubits is $(n-1)n$, whereas the number of the GMS gates with the fixed rotation angle of $\pi/2$ and arbitrary set of inputs is $2^n$, which is greater on the order than $(n-1)n$.  A more fair approach is to compare the GMS count of $12n{-}18$ to the CNOT depth of $14n{-}4$ over Linear Nearest Neighbor (LNN) architecture \cite{MR17}. This is because the number of functions computed by depth-$1$ CNOT circuits over LNN is given by the formula $\frac{2^{n+1}+(-1)^n}{3}$, and this number is similar to $2^n$.  The comparison reveals that our GMS-based construction still gives a slight advantage.

\subsection{Toffoli-$n$}
\label{SecIIIc}

We next consider the multiply-controlled gates, and specifically the multiple control Toffoli gates.  We first focus on the 3-qubit (Toffoli-3) and the 4-qubit (Toffoli-4) cases.

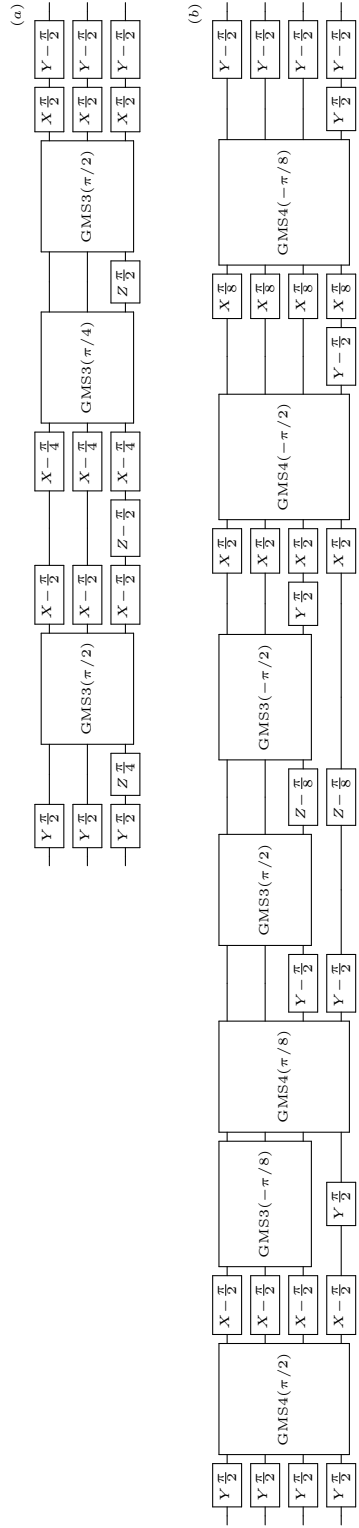
\begin{sidewaysfigure}
\tiny{
\begin{eqnarray*}
(a)
\\
\Qcircuit @C=.5em @R=.5em {
& \qw & \gate{Y\frac{\pi}{2}} & \qw                    & \multigate{2}{{\rm GMS}3(\pi/2)} & \gate{X{-}\frac{\pi}{2}} & \qw                     & \gate{X{-}\frac{\pi}{4}} & \multigate{2}{{\rm GMS}3(\pi/4)} & \qw
 & \multigate{2}{{\rm GMS}3(\pi/2)} & \gate{X\frac{\pi}{2}} & \gate{Y{-}\frac{\pi}{2}}        & \qw  & \qw \\
& \qw & \gate{Y\frac{\pi}{2}} & \qw                    & \ghost{{\rm GMS}3(\pi/2)}        & \gate{X{-}\frac{\pi}{2}} & \qw                     & \gate{X{-}\frac{\pi}{4}} &  \ghost{{\rm GMS}3(\pi/4)} & \qw 
 & \ghost{{\rm GMS}3(\pi/2)}            & \gate{X\frac{\pi}{2}} & \gate{Y{-}\frac{\pi}{2}}       & \qw  & \qw \\
& \qw & \gate{Y\frac{\pi}{2}} & \gate{Z\frac{\pi}{4}} & \ghost{{\rm GMS}3(\pi/2)}         & \gate{X{-}\frac{\pi}{2}} & \gate{Z{-}\frac{\pi}{2}}& \gate{X{-}\frac{\pi}{4}} & \ghost{{\rm GMS}3(\pi/4)} & \gate{Z\frac{\pi}{2}}
 & \ghost{{\rm GMS}3(\pi/2)}            & \gate{X\frac{\pi}{2}} & \gate{Y{-}\frac{\pi}{2}}       & \qw  & \qw 
}
\\
\\
\\
(b)
\\
\Qcircuit @C=.5em @R=.5em {
& \qw & \gate{Y\frac{\pi}{2}} & \multigate{3}{{\rm GMS}4(\pi/2)}& \gate{X{-}\frac{\pi}{2}}  & \multigate{2}{{\rm GMS}3(-\pi/8)}  
& \multigate{3}{{\rm GMS}4(\pi/8)}  &\qw                    & \multigate{2}{{\rm GMS}3(\pi/2)}  & \qw & \multigate{2}{{\rm GMS}3(-\pi/2)} 
& \qw                    & \gate{X\frac{\pi}{2}} & \multigate{3}{{\rm GMS}4(-\pi/2)} & \qw                    & \gate{X\frac{\pi}{8}} 
& \multigate{3}{{\rm GMS}4(-\pi/8)}    &\qw                  & \gate{Y{-}\frac{\pi}{2}}
& \qw & \qw \\
& \qw & \gate{Y\frac{\pi}{2}} &  \ghost{{\rm GMS}4(\pi/2)}         & \gate{X{-}\frac{\pi}{2}}  & \ghost{{\rm GMS}3(-\pi/8)} 
& \ghost{{\rm GMS}4(\pi/8)}         &\qw                    & \ghost{{\rm GMS}3(\pi/2)}           & \qw  &\ghost{{\rm GMS}3(-\pi/2)}
& \qw                    & \gate{X\frac{\pi}{2}} & \ghost{{\rm GMS}4(-\pi/2)}        & \qw                    & \gate{X\frac{\pi}{8}}
& \ghost{{\rm GMS}4(-\pi/8)}           &\qw                  & \gate{Y{-}\frac{\pi}{2}}
& \qw & \qw \\
& \qw & \gate{Y\frac{\pi}{2}} & \ghost{{\rm GMS}4(\pi/2)}          & \gate{X{-}\frac{\pi}{2}}  & \ghost{{\rm GMS}3(-\pi/8)} 
& \ghost{{\rm GMS}4(\pi/8)}         &\gate{Y{-}\frac{\pi}{2}}& \ghost{{\rm GMS}3(\pi/2)}            & \gate{Z{-}\frac{\pi}{8}} & \ghost{{\rm GMS}3(-\pi/2)}
& \gate{Y\frac{\pi}{2}} & \gate{X\frac{\pi}{2}} & \ghost{{\rm GMS}4(-\pi/2)}         & \qw                    & \gate{X\frac{\pi}{8}}
& \ghost{{\rm GMS}4(-\pi/8)}           & \qw & \gate{Y{-}\frac{\pi}{2}}
& \qw & \qw \\
& \qw & \gate{Y\frac{\pi}{2}} & \ghost{{\rm GMS}4(\pi/2)}          & \gate{X{-}\frac{\pi}{2}}  & \gate{Y\frac{\pi}{2}}
& \ghost{{\rm GMS}4(\pi/8)}        &\gate{Y{-}\frac{\pi}{2}}& \qw                                   & \gate{Z{-}\frac{\pi}{8}} & \qw
& \qw                   & \gate{X\frac{\pi}{2}} & \ghost{{\rm GMS}4(-\pi/2)}         & \gate{Y{-}\frac{\pi}{2}}& \gate{X\frac{\pi}{8}}
& \ghost{{\rm GMS}4(-\pi/8)}           &\gate{Y\frac{\pi}{2}} & \gate{Y{-}\frac{\pi}{2}}
& \qw & \qw
}
\end{eqnarray*}
}
\caption{\label{fig6}
GMS-based implementation of (a) Toffoli-3 and (b) Toffoli-4 without ancillary qubits.
}
\end{sidewaysfigure}

The efficient use of GMS gates in the case of multiply-controlled NOT (Toffoli) has previously been shown for the Toffoli-3 \cite[Figure~2]{MMNS16} and Toffoli-4 \cite[Equation~(9)]{Ivanov} gates.  In particular, reference \cite{Ivanov} presents a GMS-based circuit decomposition for the triply-controlled $Z$ gate, equivalent to the Toffoli-4 through conjugating the target by a pair of Hadamard gates. For convenience, we showed the respective constructions in Fig.~\ref{fig6} (a) and (b).  One may observe that in the case of the Toffoli-3 only 3 GMS gates are needed, compared to 5 local two-qubit gates \cite{NC, FMLLDM17} and, for the Toffoli-4, only 7 GMS gates are needed, compared to 11 local two-qubit gates \cite{FMLLDM17}. We note that, unlike in \cite{FMLLDM17}, the 7-GMS Toffoli-4 construction of \cite{Ivanov} furthermore does not require an ancillary qubit.

In pursuit of further gate count reduction, we consider employing ancillary qubits in our GMS-based construction of the $n$-qubit Toffoli gate. The employment of ancillary qubits to reduce the gate counts in constructing the Toffoli-$n$ gate has in fact been extensively investigated in \cite{BBCD96, M16}, but in the context of relying on the local entangling gates. Using ancillae turns out to be helpful in the case of quantum circuits employing the GMS gate, as well.  In the following, we show a step-by-step construction of the GMS-based ancilla-aided Toffoli-4 gate (we report no improvements to the Toffoli-3 circuit).

We start with a simple observation that the Toffoli-4 gate is equivalent to the CCCZ gate up to the conjugation by the Hadamard gates, such as illustrated next,
\[
\Qcircuit @C=.1em @R=.1em @!{
& \qw      & \ctrl{1} & \qw      & \qw \\
& \qw      & \ctrl{1} & \qw      & \qw \\
& \qw      & \ctrl{1} & \qw      & \qw \\
& \gate{H} & \gate{Z} & \gate{H} & \qw 
}
\raisebox{-2.4em}{\hspace{1mm}=\hspace{1mm}}
\Qcircuit @C=.7em @R=.8em @!R {
& \ctrl{1} 	& \qw \\
& \ctrl{1} 	& \qw \\
& \ctrl{1} 	& \qw \\
& \targ 	& \qw 	
}
\]
CCCZ$(a,b,c,d)$ gate performs the transformation 
$$\ket{abcd}\mapsto (-1)^{abcd}\ket{abcd} = (e^{i\pi/8})^{8abcd}\ket{abcd} = w_{16}^{8abcd}\ket{abcd},$$ 
where $w_{16}$ is the primitive $16^{\text{th}}$ complex root of the number $1$.  Using mixed arithmetic equality $2xy=x + y - (x \oplus y)$ three times allows to rewrite the above formula as $$\ket{abcd} \mapsto w_{16}^{a+b+c+d-(a{\oplus}b)-(a \oplus c)-(a \oplus d)-(b \oplus c)-(b \oplus d)-(c \oplus d)+(a \oplus b \oplus c)+(a \oplus b \oplus d)+(a \oplus c \oplus d)+(b \oplus c \oplus d)-(a \oplus b \oplus c \oplus d)}\ket{abcd}.$$  
This function can thus be implemented as a CNOT and $\text{RZ}(\pm\frac{\pi}{8})$ circuit by applying $Z$ rotation with the positive sign to the linear terms $\{a,b,c,d,a{\oplus}b{\oplus}c,a{\oplus}b{\oplus}d, a{\oplus}c{\oplus}d, b{\oplus}c{\oplus}d\}$ and $Z$ rotation with the negative sign to the terms $\{a{\oplus}b, a{\oplus}c, a{\oplus}d, b{\oplus}c, b{\oplus}d, c{\oplus}d, a{\oplus}b{\oplus}c{\oplus}d\}$, with each such linear term obtainable by the CNOT gates.  In the next, we will show how to induce all necessary CNOT gates to allow the application of the necessary RZ gates, using only a few GMS gates. 

We first note that the linear functions with the single literate each, $\{a,b,c,d\}$, are the original qubits provided to us on the input side of the circuit.  Therefore, all length-1 linear terms may be implemented by simply applying $\text{RZ}(\frac{\pi}{8})$ single-qubit rotation gates to each respective qubit.  By doing so, we construct the circuit
\[
\Qcircuit @C=.1em @R=.1em @!{
\lstick{a} & \gate{Z\frac{\pi}{8}} & \qw \\
\lstick{b} & \gate{Z\frac{\pi}{8}} & \qw \\
\lstick{c} & \gate{Z\frac{\pi}{8}} & \qw \\
\lstick{d} & \gate{Z\frac{\pi}{8}} & \qw 	
}
\]
using no GMS gates and implementing the transformation $\ket{abcd}\mapsto w_{16}^{a+b+c+d} \ket{abcd}$.  We next have to find how to apply as few as possible GMS gates in a way that enables to exercise the remaining $11$ $Z$ rotations.  

\begin{figure}
\begin{eqnarray*}
\Qcircuit @C=.8em @R=.3em @!R {
\lstick{a}		& \gate{Z\frac{\pi}{8}} & \ctrl{4}& \qw     & \qw     	& \qw     & \qw              			& \qw     & \qw     & \qw     & \ctrl{4} & \qw \\
\lstick{b}		& \gate{Z\frac{\pi}{8}} & \qw     & \ctrl{3}& \qw    	& \qw     & \qw              			& \qw     & \qw     & \ctrl{3} & \qw     & \qw \\
\lstick{c}		& \gate{Z\frac{\pi}{8}} & \qw     & \qw     & \ctrl{2}	& \qw     & \qw             			& \qw     & \ctrl{2} & \qw     & \qw     & \qw \\
\lstick{d}		& \gate{Z\frac{\pi}{8}} & \qw     & \qw     & \qw    	& \ctrl{1}& \qw              			& \ctrl{1} & \qw     & \qw     & \qw     & \qw \\
\lstick{\ket{0}}& \qw 					& \targ   & \targ   & \targ		& \targ   & \gate{Z{-}\frac{\pi}{8}}	& \targ    & \targ  & \targ   & \targ    & \qw
}\\
\raisebox{-4em}{=\hspace{7mm}}
\Qcircuit @C=.3em @R=.3em {
\lstick{a}& \gate{Z\frac{\pi}{8}} & \gate{Y\frac{\pi}{2}}          & \multigate{4}{{\rm GMS}5(\pi/2)}  & \qw    & \qw                       & \qw  & \multigate{4}{{\rm GMS}5(-\pi/2)} & \qw & \gate{Y{-}\frac{\pi}{2}}        & \qw  & \qw & \qw \\
\lstick{b}& \gate{Z\frac{\pi}{8}} & \gate{Y\frac{\pi}{2}}          &  \ghost{{\rm GMS}5(\pi/2)}            & \qw   & \qw                       & \qw  & \ghost{{\rm GMS}5(-\pi/2)}            & \qw & \gate{Y{-}\frac{\pi}{2}}       & \qw  & \qw & \qw \\
\lstick{c}& \gate{Z\frac{\pi}{8}} & \gate{Y\frac{\pi}{2}}          & \ghost{{\rm GMS}5(\pi/2)}             & \qw   & \qw                       & \qw  & \ghost{{\rm GMS}5(-\pi/2)}            & \qw & \gate{Y{-}\frac{\pi}{2}}       & \qw  & \qw & \qw \\
\lstick{d}& \gate{Z\frac{\pi}{8}} & \gate{Y\frac{\pi}{2}}          & \ghost{{\rm GMS}5(\pi/2)}             & \qw   & \qw                       & \qw  & \ghost{{\rm GMS}5(-\pi/2)}            & \qw & \gate{Y{-}\frac{\pi}{2}}       & \qw  & \qw & \qw \\
\lstick{\ket{0}}& \qw & \qw                             & \ghost{{\rm GMS}5(\pi/2)}             & \qw   & \gate{Z{-}\frac{\pi}{8}}        & \qw  & \ghost{{\rm GMS}5(-\pi/2)}            & \qw & \qw                            & \qw  & \qw & \qw 
}
\end{eqnarray*}
\caption{\label{figabcd} Obtaining phase $w_{16}^{-(a \oplus b \oplus c \oplus d)}$.}
\end{figure}
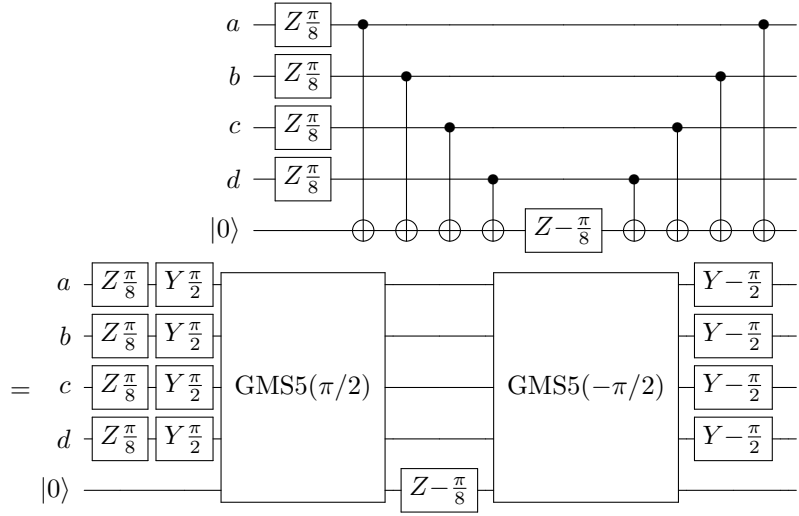

To apply the $Z$ rotation to the length-4 linear term, $a{\oplus}b{\oplus}c{\oplus}d$, we introduce an ancillary qubit in the $\ket{0}$ state, copy all qubits into it using a set of four CNOTs sharing the target, and then uncompute those CNOTs.  This allows to apply one new $Z$ rotation between the two layers of the CNOT gates, and the number of the GMS gates required to implement this construction is two, see Fig.~\ref{figabcd}.
The circuit constructed thus far performs the transformation $\ket{abcd}\mapsto w_{16}^{a+b+c+d-(a \oplus b \oplus c \oplus d)} \ket{abcd}$. Observe that each of the two sets of the CNOT gates on the left hand side of the circuit equality in Fig.~\ref{figabcd} requires two GMS gates to be implemented (both are fan-in gates, considered earlier), for a total of four GMS gates, two GMS5 and two GMS4.  However, the two GMS4 can be chosen with the opposite signs and they commute with all other gates we are about to introduce in the middle to cancel out.  This means that only two GMS5 gates are needed in our construction.  

We next need to apply the remaining 10 $Z$ rotations to obtain the desired CCCZ gate.  To do so, consider the following circuit identity:
\[
\Qcircuit @C=.7em @R=1em {
&\lstick{x} & \ctrl{1} & \qw            & \ctrl{1}& \qw \\
&\lstick{y} & \targ    & \gate{Z\theta} & \targ   & \qw 
}
\raisebox{-1em}{\hspace{2mm}$\equiv$\hspace{2mm}}
\Qcircuit @C=.7em @R=.5em {
& \gate{H} & \multigate{1}{{\rm XX}(\theta)} 	& \gate{H} 	& \qw \\
& \gate{H} & \ghost{{\rm XX}(\theta)}           & \gate{H} 	& \qw 
}
\]
where the left hand side, trivially, performs a phase rotation by the angle $\theta$ applied to the linear function $x \oplus y$ and the right hand side reports an equivalent circuit based on the XX gate, up to a global phase.  We can generalize this construction to $n$ qubits, by replacing the XX gate with the GMS on the right hand side, while conjugating by the layer of Hamadards before and after.  What this accomplishes is the application of phases to EXORs of all pairs of participating variables, as described by the circuit on the left hand side.  Formally, 
\begin{eqnarray*}
H[x_1]H[x_2]...H[x_n]\text{GMS}n[x_1,x_2,...,x_n](\theta)H[x_1]H[x_2]...H[x_n]: \\ 
\ket{x_1x_2...x_n} \mapsto e^{i\theta\sum_{j<k}x_j\oplus x_k}\ket{x_1x_2...x_n}.
\end{eqnarray*}

\begin{figure}[t]
\begin{eqnarray*}
\Qcircuit @C=.3em @R=.62em @!R{
\lstick{a} 		& \ctrl{1} 	& \qw \\
\lstick{b} 		& \ctrl{1} 	& \qw \\
\lstick{c} 		& \ctrl{1} 	& \qw \\
\lstick{d} 		& \gate{Z} 	& \qw \\
\lstick{\ket{0}}& \qw 		& \qw & \rstick{\ket{0}}
}
\raisebox{-3.75em}{\hspace{7mm}=\hspace{7mm}}
\Qcircuit @C=.25em @R=.25em {
\lstick{a}& \gate{Z\frac{\pi}{8}}& \qw & \ctrl{4} & \qw     & \qw     & \qw     & \qw & \qw &\gate{H} & \qw & \multigate{4}{{\rm GMS}5(\pi/8)} & \qw & \multigate{3}{{\rm GMS}4(-\pi/4)} & \qw         &\gate{H} & \qw & \qw     & \qw     & \qw     & \ctrl{4} & \qw \\
\lstick{b}& \gate{Z\frac{\pi}{8}}& \qw & \qw     & \ctrl{3} & \qw    & \qw     & \qw & \qw &\gate{H}  & \qw  & \ghost{{\rm GMS}5(\pi/8)}          & \qw & \ghost{{\rm GMS}4(-\pi/4)} & \qw       &\gate{H} & \qw & \qw     & \qw     & \ctrl{3} & \qw     & \qw \\
\lstick{c}& \gate{Z\frac{\pi}{8}}& \qw & \qw     & \qw     & \ctrl{2}& \qw     & \qw & \qw &\gate{H}  & \qw  & \ghost{{\rm GMS}5(\pi/8)}          & \qw & \ghost{{\rm GMS}4(-\pi/4)}& \qw       &\gate{H} & \qw & \qw     & \ctrl{2} & \qw     & \qw      & \qw \\
\lstick{d}& \gate{Z\frac{\pi}{8}}& \qw & \qw     & \qw      & \qw    & \ctrl{1}& \qw & \qw &\gate{H}  & \qw  & \ghost{{\rm GMS}5(\pi/8)}          & \qw & \ghost{{\rm GMS}4(-\pi/4)}& \qw       &\gate{H} & \qw & \ctrl{1} & \qw     & \qw     & \qw     & \qw \\
\lstick{\ket{0}} & \qw & \qw & \targ   & \targ    & \targ     & \targ  & \qw & \gate{Z{-}\frac{\pi}{8}} &\gate{H}  & \qw  & \ghost{{\rm GMS}5(\pi/8)}         & \qw & \qw & \qw      &\gate{H} & \qw & \targ    & \targ  & \targ   & \targ    & \qw & \rstick{\ket{0}}
} \\
\raisebox{-3.55em}{=\hspace{5mm}}
\small{
\Qcircuit @C=.3em @R=.3em @!R {
\lstick{a}& \gate{Z\frac{\pi}{8}} & \gate{Y\frac{\pi}{2}} & \multigate{4}{{\rm GMS}5(\pi/2)} & \qw 
& \qw                    & \multigate{4}{{\rm GMS}5(\pi/8)} & \multigate{3}{{\rm GMS}4(-\pi/4)} & \multigate{4}{{\rm GMS}5(-\pi/2)} & \gate{Y{-}\frac{\pi}{2}} & \qw &\qw\\
\lstick{b}& \gate{Z\frac{\pi}{8}} & \gate{Y\frac{\pi}{2}} & \ghost{{\rm GMS}5(\pi/2)}           & \qw            
& \qw                    & \ghost{{\rm GMS}5(\pi/8)}           & \ghost{{\rm GMS}4(-\pi/4)}     & \ghost{{\rm GMS}5(-\pi/2)}           & \gate{Y{-}\frac{\pi}{2}} &\qw &\qw\\
\lstick{c}& \gate{Z\frac{\pi}{8}} & \gate{Y\frac{\pi}{2}} & \ghost{{\rm GMS}5(\pi/2)}           & \qw            
& \qw                    & \ghost{{\rm GMS}5(\pi/8)}          & \ghost{{\rm GMS}4(-\pi/4)}      & \ghost{{\rm GMS}5(-\pi/2)}           & \gate{Y{-}\frac{\pi}{2}} &\qw &\qw\\
\lstick{d}& \gate{H} & \gate{X\frac{\pi}{8}}                    & \ghost{{\rm GMS}5(\pi/2)}     & \qw            
& \qw                    & \ghost{{\rm GMS}5(\pi/8)}          & \ghost{{\rm GMS}4(-\pi/4)}      & \ghost{{\rm GMS}5(-\pi/2)}           & \gate{H} &\qw &\qw\\
\lstick{\ket{0}} & \qw & \qw                                 & \ghost{{\rm GMS}5(\pi/2)}        & \gate{Z{-}\frac{\pi}{8}}                            
& \gate{Y\frac{\pi}{2}} & \ghost{{\rm GMS}5(\pi/8)}          & \gate{Y{-}\frac{\pi}{2}} & \ghost{{\rm GMS}5(-\pi/2)}           & \qw                   
     &\qw& \qw & \rstick{\ket{0}}
}
}
\end{eqnarray*}
\caption{\label{figCCCZ} CCCZ gate using four GMS gates. Right hand side shows two Hadamard gates, removing which transforms CCCZ gate into Toffoli-4.}
\end{figure}
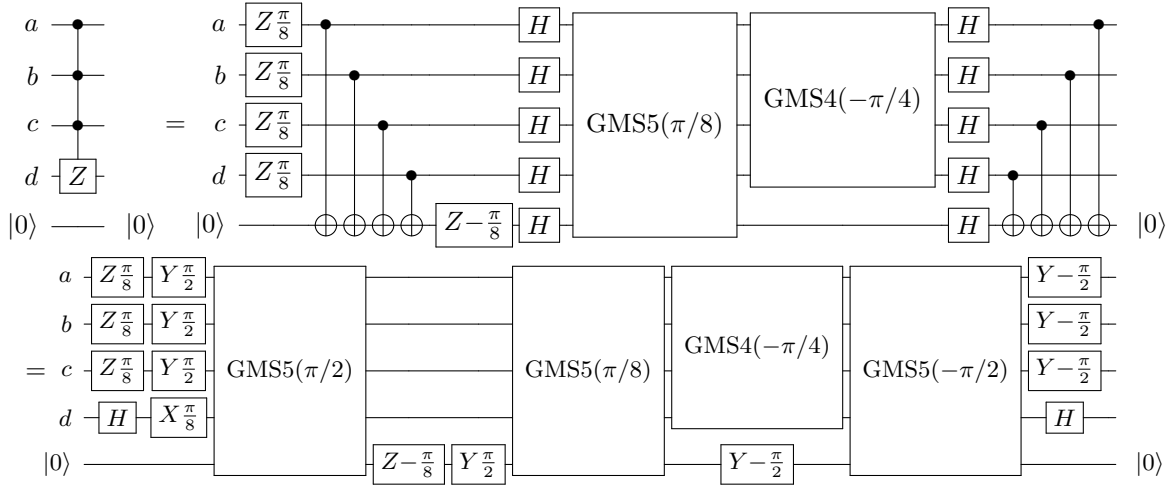

We next apply the above identity over GMS to our ongoing construction of the Toffoli-4 gate. To obtain length-3 linear functions, we may insert Hadamard-conjugated GMS5$(\pi/8)$ in the middle of our current circuit (Fig.\ref{figabcd}).  The effect this has is the introduction of phase $\frac{\pi}{8}$ applied to all pairs of qubits participating in the construction. In the middle of the circuit the qubits we have are described by the linear functions $\{a,b,c,d,a{\oplus}b{\oplus}c{\oplus}d)\}$.  Thus, the set of EXOR pairs is $\{a{\oplus}b, a{\oplus}c, a{\oplus}d, b{\oplus}c, b{\oplus}d, c{\oplus}d, a{\oplus}b{\oplus}c, a{\oplus}b{\oplus}d, a{\oplus}c{\oplus}d, b{\oplus}c{\oplus}d\}$. This means that the overall action preformed by the circuit with 3 GMS gates can be written as 
$$\ket{abcd} \mapsto w_{16}^{a+b+c+d-(a \oplus b \oplus c \oplus d) \;+\; (a{\oplus}b)+(a \oplus c)+(a \oplus d)+(b \oplus c)+(b \oplus d)+(c \oplus d)+(a \oplus b \oplus c)+(a \oplus b \oplus d)+(a \oplus c \oplus d)+(b \oplus c \oplus d)}\ket{abcd}.$$ 
Observe that the signs of the length-2 terms are not the ones we wanted to have.  This may, however, be corrected by applying Hadamard-conjugated GMS4$(-\pi/4)$ to the qubits $\{a,b,c,d\}$, resulting in the phase correction by the product $w_{16}^{-2(a{\oplus}b)-2(a \oplus c)-2(a \oplus d)-2(b \oplus c)-2(b \oplus d)-2(c \oplus d)}$, and leading to the desired transformation 
$$\ket{abcd} \mapsto w_{16}^{a+b+c+d-(a{\oplus}b)-(a \oplus c)-(a \oplus d)-(b \oplus c)-(b \oplus d)-(c \oplus d)+(a \oplus b \oplus c)+(a \oplus b \oplus d)+(a \oplus c \oplus d)+(b \oplus c \oplus d)-(a \oplus b \oplus c \oplus d)}\ket{abcd},$$ 
accomplished as a 4-GMS circuit shown in Fig.~\ref{figCCCZ}.

Using a similar approach, we can obtain a 3-GMS circuit implementing the CCZ gate on qubits $a$, $b$, and $c$, as follows:
\[
\Qcircuit @C=.5em @R=.5em @!R {
\lstick{a} & \qw     & \gate{Z\frac{\pi}{4}} & \gate{Y\frac{\pi}{2}} & \multigate{3}{{\rm GMS}4(\pi/2)} & \multigate{2}{{\rm GMS}3(-\pi/4)}       & \multigate{3}{{\rm GMS}4(-\pi/2)} & \gate{Y{-}\frac{\pi}{2}} & \qw \\
\lstick{b} & \qw     & \gate{Z\frac{\pi}{4}} & \gate{Y\frac{\pi}{2}} & \ghost{{\rm GMS}4(\pi/2)}           & \ghost{{\rm GMS}3(-\pi/4)}                  & \ghost{{\rm GMS}4(-\pi/2)}           & \gate{Y{-}\frac{\pi}{2}} &\qw \\
\lstick{c} & \qw     & \gate{H}                  & \gate{X\frac{\pi}{4}} & \ghost{{\rm GMS}4(\pi/2)}           & \ghost{{\rm GMS}3(-\pi/4)}                  & \ghost{{\rm GMS}4(-\pi/2)}           & \gate{H} &\qw \\
\lstick{\ket{0}} & \qw & \qw & \qw                                    & \ghost{{\rm GMS}4(\pi/2)}           & \gate{Z\frac{\pi}{4}}                                & \ghost{{\rm GMS}4(-\pi/2)}           & \qw &\qw & \rstick{\ket{0}}
}
\]
It is different from those reported in \cite{MMNS16,MKH09}.

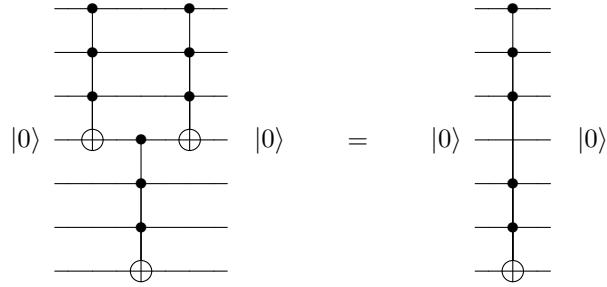
\begin{figure}
\begin{eqnarray*}
\Qcircuit @C=.5em @R=.83em @!R {
			& \qw & \ctrl{3} & \qw & \qw     & \qw  & \ctrl{3} & \qw & \qw & \\
			& \qw & \ctrl{2} & \qw & \qw     & \qw  & \ctrl{2} & \qw & \qw & \\
			& \qw & \ctrl{1} & \qw & \qw     & \qw  & \ctrl{1} & \qw & \qw & \\
\lstick{\ket{0}}    & \qw & \targ   & \qw & \ctrl{3} & \qw  & \targ    & \qw & \qw & \rstick{\ket{0}} \\
			& \qw & \qw     & \qw & \ctrl{2} & \qw  & \qw      & \qw & \qw & \\
			& \qw & \qw     & \qw & \ctrl{1} & \qw  & \qw      & \qw & \qw & \\
			& \qw & \qw     & \qw & \targ    & \qw  & \qw      & \qw & \qw & 
}
\qquad\qquad
\raisebox{-5em}{=}
\qquad\qquad
\Qcircuit @C=.5em @R=.83em @!R {
			&\qw &\ctrl{6} &\qw &\qw & \\
			&\qw &\ctrl{5} &\qw &\qw & \\
			&\qw &\ctrl{4} &\qw &\qw & \\
\lstick{\ket{0}}    &\qw &\qw     &\qw &\qw & \rstick{\ket{0}} \\ 
			&\qw &\ctrl{2} &\qw &\qw & \\
			&\qw &\ctrl{1} &\qw &\qw & \\
			&\qw &\targ    &\qw &\qw &
}
\end{eqnarray*}
\caption{\label{fig8}
Ancilla-aided construction of the Toffoli-6 using a set of three Toffoli-4, each of which is constructible using three GMS gates.
}
\end{figure}

In the remaining part of this subsection, we briefly outline an implementation of the $n$-qubit Toffoli gate using $3n{-}9$ GMS gates and $\frac{n-2}{2}$ ancillae for even $n$, and $3n{-}6$ GMS gates and $\frac{n-1}{2}$ ancillae for odd $n$, $n \geq 6$.  This beats $6n{-}12$ local CNOT gates result of \cite{M16}, while using a comparable number of ancillae. Our construction relies on nesting efficient 3-GMS Toffoli-4 gates (shown in Fig.~\ref{figCCCZ3}; we describe how to get to this construction in the next section), such as illustrated in Fig.~\ref{fig8}, to obtain larger multiple control Toffoli gates. For odd $n$, one pair of 3-GMS Toffoli-3 gates needs to be used (equivalently, a set of two relative-phase Toffoli-3 gates, requiring 3 local entangling operations each \cite{M16}), explaining the difference between gate counts for odd and even $n$.

\section{GMS with other parameters}
\label{SecIV}

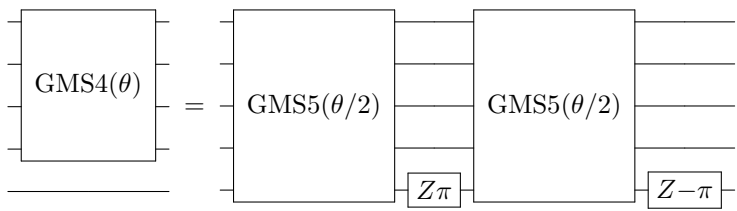
\begin{figure}[t]
\begin{eqnarray*}
\Qcircuit @C=.5em @R=.7em @!R{
& \multigate{3}{{\rm GMS}4(\theta)} & \qw \\
& \ghost{{\rm GMS}4(\theta)}        & \qw \\
& \ghost{{\rm GMS}4(\theta)}        & \qw \\
& \ghost{{\rm GMS}4(\theta)}        & \qw \\
& \qw                                & \qw \\
}
\raisebox{-3.2em}{\hspace{2mm}=\hspace{2mm}}
\Qcircuit @C=.5em @R=.22em @!R {
& \multigate{4}{{\rm GMS}5(\theta/2)} 	& \qw 			& \multigate{4}{{\rm GMS}5(\theta/2)} 	& \qw 				& \qw \\
& \ghost{{\rm GMS}5(\theta/2)}          & \qw 			& \ghost{{\rm GMS}5(\theta/2)}       	& \qw 				& \qw \\
& \ghost{{\rm GMS}5(\theta/2)}          & \qw 			& \ghost{{\rm GMS}5(\theta/2)}       	& \qw 				& \qw \\
& \ghost{{\rm GMS}5(\theta/2)}          & \qw 			& \ghost{{\rm GMS}5(\theta/2)}       	& \qw 				& \qw \\
& \ghost{{\rm GMS}5(\theta/2)}          & \gate{Z\pi}	& \ghost{{\rm GMS}5(\theta/2)}       	& \gate{Z{-}\pi} & \qw \\
}
\end{eqnarray*}
\caption{\label{fig12}
GMS$(n{-}1)$ using two GMS$n$, illustrated in the case of $n=5$.
}
\end{figure}

\begin{figure}[t]
	\begin{eqnarray*}
		\hspace{6mm}
		\small{
			\Qcircuit @C=.3em @R=.65em @!R{
				\lstick{a} 		& \ctrl{1} 	& \qw \\
				\lstick{b} 		& \ctrl{1} 	& \qw \\
				\lstick{c} 		& \ctrl{1} 	& \qw \\
				\lstick{d} 		& \gate{Z} 	& \qw \\
				\lstick{\ket{0}\hspace{-1mm}}& \qw 		& \qw & \rstick{\hspace{-2mm}\ket{0}}
			}
		}
		\raisebox{-3.8em}{\hspace{3mm}$\equiv$\hspace{5mm}}
		\small{
			\Qcircuit @C=.3em @R=.3em @!R {
				\lstick{a}& \gate{Z\frac{\pi}{8}} & \gate{Y\frac{\pi}{2}} & \multigate{4}{{\rm GMS}5(\pi/2)}
				& \qw                    & \qw & \multigate{4}{{\rm GMS}5(-\pi/8)}   & \qw & \multigate{4}{{\rm GMS}5(-\pi/2)} & \gate{Y{-}\frac{\pi}{2}} & \qw &\qw\\
				\lstick{b}& \gate{Z\frac{\pi}{8}} & \gate{Y\frac{\pi}{2}} & \ghost{{\rm GMS}5(\pi/2)}           
				& \qw                    & \qw & \ghost{{\rm GMS}5(-\pi/8)}          & \qw     & \ghost{{\rm GMS}5(-\pi/2)}           & \gate{Y{-}\frac{\pi}{2}} &\qw &\qw\\
				\lstick{c}& \gate{Z\frac{\pi}{8}} & \gate{Y\frac{\pi}{2}} & \ghost{{\rm GMS}5(\pi/2)}           
				& \qw                    & \qw & \ghost{{\rm GMS}5(-\pi/8)}          & \qw      & \ghost{{\rm GMS}5(-\pi/2)}           & \gate{Y{-}\frac{\pi}{2}} &\qw &\qw\\
				\lstick{d}& \gate{H} & \gate{X\frac{\pi}{8}}                    & \ghost{{\rm GMS}5(\pi/2)}      
				& \qw                    & \qw & \ghost{{\rm GMS}5(-\pi/8)}          & \qw      & \ghost{{\rm GMS}5(-\pi/2)}           & \gate{H} &\qw &\qw\\
				\lstick{\ket{0}\hspace{-1mm}} & \qw & \qw                                 & \ghost{{\rm GMS}5(\pi/2)} 
				& \gate{Y{-}\frac{\pi}{2}} & \gate{X\frac{\pi}{8}} & \ghost{{\rm GMS}5(-\pi/8)}         &  \gate{Y\frac{\pi}{2}} & \ghost{{\rm GMS}5(-\pi/2)}           & \qw                   
				&\qw& \qw & \rstick{\hspace{-2mm}\ket{0}}
			}
		}
	\end{eqnarray*}
	\caption{\label{figCCCZ3} Optimized implementation of the CCCZ gate using three GMS gates. Right hand side shows two Hadamard gates, removing which transforms CCCZ gate into Toffoli-4.}
\end{figure}
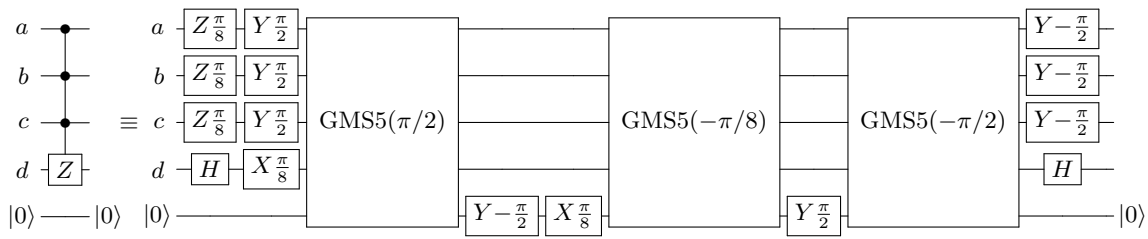

So far, we focused on using GMS gates with all equal rotation angles $\chi_{ij}$, and an arbitrarily selectable subset of qubits those global gates apply to. Such gates may not always be possible to obtain directly in an experiment.  Indeed, one possible experimental setup \cite{MMNS16} allows for the application of global Molmer-Sorensen gates affecting all $n$ qubits participating in the computation.  As such, an $(n{-}1)$-qubit GMS gate may not be directly available on an $n$-qubit system. To circumvent this and enable smaller GMS gates, we propose the following. First, start with the circuit identity
\[
\Qcircuit @C=.5em @R=.7em @!R {
& \multigate{1}{{\rm XX}(\theta)} & \qw               & \multigate{1}{{\rm XX}(\theta)}& \qw \\
& \ghost{{\rm XX}(\theta)}           & \gate{Z\pi}& \ghost{{\rm XX}(\theta)}           & \qw 
}
\raisebox{-1em}{\hspace{2mm}=\hspace{2mm}}
\Qcircuit @C=.5em @R=.7em @!R {
& \qw                & \qw \\
& \gate{Z\pi}& \qw 
}
\]
Using the identity recursively, such as illustrated next,
\[
\Qcircuit @C=.4em @R=.7em @!R {
& \multigate{2}{{\rm XX}(\theta_2)} & \qw                               & \qw           & \qw       						& \multigate{2}{{\rm XX}(\theta_2)} & \qw \\
& \ghost{{\rm XX}(\theta_2)}        & \multigate{1}{{\rm XX}(\theta_1)} & \qw           & \multigate{1}{{\rm XX}(\theta_1)} & \ghost{{\rm XX}(\theta_2)} 		& \qw \\
& \ghost{{\rm XX}(\theta_2)}        & \ghost{{\rm XX}(\theta_1)}   		& \gate{Z\pi}	& \ghost{{\rm XX}(\theta_1)}   		& \ghost{{\rm XX}(\theta_2)}     	& \qw \\
}
\raisebox{-2.1em}{\hspace{2mm}=\hspace{2mm}}
\Qcircuit @C=.4em @R=.7em @!R {
& \multigate{2}{{\rm XX}(\theta_2)} 	& \qw               & \multigate{2}{{\rm XX}(\theta_2)}  	& \qw \\
& \ghost{{\rm XX}(\theta_2)}           	& \qw               & \ghost{{\rm XX}(\theta_2)}            & \qw \\
& \ghost{{\rm XX}(\theta_2)}           	& \gate{Z\pi}		& \ghost{{\rm XX}(\theta_2)}           	& \qw \\
}
\raisebox{-2.1em}{\hspace{2mm}=\hspace{2mm}}
\Qcircuit @C=.5em @R=.7em @!R {
& \qw        	& \qw \\
& \qw         	& \qw \\
& \gate{Z\pi}	& \qw \\
}
\]
we arrive at the conclusion that the ${\rm RZ}(\pi)$ gate effects a spin echo on the identical XX gates to its left and right, provided that the qubit that the ${\rm RZ}(\pi)$ applies to also participates in the XX gates, and as a result cancels out the respective XX interactions.  Based on this property, Fig.~\ref{fig12} illustrates how to obtain an $(n{-}1)$-qubit GMS gate out of two $n$-qubit GMS gates.  The construction can be used iteratively to obtain global gates spanning arbitrarily selectable subsets of qubits, and enabling all constructions described in Section \ref{SecIII} in the case when only the maximal size GMS gate is available.  In fact, this inspired the construction of a more efficient Toffoli-4 implementation.  Specifically, Toffoli-4 gate may be obtained using only $3$ maximal size GMS gates on a 5-qubit machine.  This is because substituting $\text{GMS}4(-\pi/4)$ (Fig.~\ref{fig12}) into GMS-enabled implementation of the Toffoli-4 gate (Fig.~\ref{figCCCZ}) results in the circuit over $5$ GMS5 gates, however, $\text{GMS}5(\pi/8)$ used in Fig.~\ref{figCCCZ} meets the newly introduced $\text{GMS}5(-\pi/8)$ and they cancel out, reducing the GMS gate count to $3$.  This improved construction is illustrated in Fig.~\ref{figCCCZ3}. Our optimized 3-GMS Toffoli-4 construction relies on notably fewer entangling pulses compared to $11$ two-qubit gates in \cite{FMLLDM17} or $7$ GMS gates in \cite{Ivanov}.

\vspace{1mm}
The signs of $\chi_{ij}$ may furthermore be determined by the experiment such as the case in \cite{DLF16}, disallowing their uniform assignment.  It is, however, expected \cite{Monroe-PC} that future trapped ions experiments will feature a fully controllable sign of the interaction, and this will not be an issue.  Should the signs be uncontrollable, this provides an additional challenge, since the constructions in Fig.~\ref{fig1} and Fig.~\ref{fig2} rely on the ability to apply GMS gates with the inverted sign of the rotation angle.  In case when the signs cannot be controlled individually, the inverse GMS gate can, in fact, be induced by the single-qubit corrections applied to the GMS gates with uncontrollable parameter signs as follows.

First, start with the following identity
\begin{equation*}
\text{XX}_{ij}^{\dagger}(\chi) := \text{XX}_{ij}(-\chi)
= -i\cdot {\rm RX}_i(\pi){\rm RX}_j(\pi)\text{XX}_{ij}(\pi-\chi),
\end{equation*}
where ${\rm RX}_k$ denotes the RX gate applied to the $k^{\text{th}}$ qubit.
Using this identity $\frac{n(n{-}1)}{2}$ times allows to construct the $\text{GMS}^\dagger$ using only one GMS gate, as follows: 
\begin{align}
\text{GMS}^{\dagger} (\chi)  := \prod_{i=1}^n \prod_{j=i+1}^{n} \text{XX}_{ij}(-\chi) \nonumber \\ 
= (-i)^{n(n-1)/2} \prod_{i=1}^n \prod_{j=i+1}^{n}
\text{RX}_i(\pi) \text{RX}_j(\pi)\text{XX}_{ij}(\pi-\chi) \nonumber \\
= (-i)^{n(n-1)/2} \left(\prod_{i=1}^n \text{RX}_i(\pi)\right)^{n-1}
\left(\prod_{i=1}^n \prod_{j=i+1}^{n}\text{XX}_{ij}(\pi-\chi)\right) \nonumber \\
= (-i)^{n(n-1)/2} \left(\prod_{i=1}^n \text{RX}_i((n{-}1)\pi)\right) \text{GMS}(\pi{-}\chi). \nonumber 
\end{align}

In other words, whenever GMS$^\dagger$ is not directly available due to the inability to invert the sign of the interactions, {\em i.e.}, $\chi \in [0,\pi]$, the GMS$^\dagger$ may be still be constructed with the use of a single GMS gate by taking the parameter value of $(\pi-\chi) \in [0,\pi]$, and performing single-qubit corrections.  This enables constructions from Figs.~\ref{fig1} and \ref{fig2} in the scenario with uncontrollable signs of the individual interactions within GMS.

\subsection{Quantum Fourier Arithmetic}\label{SecIIId}

Previously, we considered the case where all $|\chi_{ij}|$ are constant, regardless of the choice of $i$ and $j$.  However, it is possible that $|\chi_{ij}|$ drops off as a function of the distance $|i-j|$.  This may be natural given physical interaction strengths typically scale in the distance between qubits \cite{Hess}.

\begin{figure}
\begin{eqnarray*}
(a)
\\
\Qcircuit @C=.5em @R=.95em {
& \qw & \multigate{4}{QFT} & \qw & \qw \\
& \qw & \ghost{QFT}           & \qw & \qw \\
& \qw & \ghost{QFT}           & \qw & \qw \\
& \qw & \ghost{QFT}           & \qw & \qw \\
& \qw & \ghost{QFT}           & \qw & \qw
}
\raisebox{-3.8em}{\hspace{1mm}=\hspace{1mm}}
\Qcircuit @C=.5em @R=.4em @!R {
& \qw     & \qw & \gate{\theta_4}          & \qw       & \gate{\theta_3}         & \qw       & \gate{\theta_2}         &\qw        &\gate{\theta_1} &\gate{H} &\qw&\qw\\
& \qw     & \qw & \gate{\theta_3} \qwx & \qw       & \gate{\theta_2} \qwx & \qw       & \gate{\theta_1} \qwx &\gate{H} &\ctrl{-1}             &\qw        &\qw&\qw\\
& \qw     & \qw & \gate{\theta_2} \qwx & \qw       & \gate{\theta_1} \qwx & \gate{H}&\ctrl{-1}                      &\qw        &\qw                   &\qw        &\qw&\qw\\ 
& \qw     & \qw & \gate{\theta_1} \qwx & \gate{H}& \ctrl{-1}                     & \qw       &\qw                            &\qw        &\qw                   &\qw        &\qw&\qw\\
& \qw     & \gate{H} & \ctrl{-1}              & \qw       & \qw                           & \qw       &\qw                            &\qw        &\qw                   &\qw        &\qw&\qw
}
\end{eqnarray*}
\begin{eqnarray*}
\\
(b)
\\
\Qcircuit @C=.5em @R=1.37em @!R{
\lstick{\ket{a}} & \qw  & \ctrl{1}&\qw & \qw & \\
& \qw & \multigate{4}{QFA} & \qw & \qw \\
& \qw & \ghost{QFA}           & \qw & \qw \\
& \qw & \ghost{QFA}           & \qw & \qw \\
& \qw & \ghost{QFA}           & \qw & \qw \\
& \qw & \ghost{QFA}           & \qw & \qw
}
\raisebox{-5.7em}{\hspace{1mm}=\hspace{7mm}}
\Qcircuit @C=.5em @R=.43em @!R {
\lstick{\ket{a}} &\qw   &\qw             & \qw & \ctrl{1}                             & \qw & \ctrl{2}                            & \qw & \ctrl{3}                             & \qw & 
\ctrl{4}                             &\qw & \ctrl{5} &\qw & \qw & \qw & \qw \\
& \qw & \multigate{4}{QFT} & \qw & \gate{\theta_0^{a_{[0]}}}         & \qw & \qw                                 & \qw & \qw                                  & \qw &
\qw                                 & \qw & \qw                                  & \qw &\multigate{4}{QFT^\dagger} & \qw & \qw \\
& \qw & \ghost{QFT}           & \qw & \gate{\theta_1^{a_{[0]}}} \qwx & \qw & \gate{\theta_0^{a_{[1]}}}         & \qw & \qw                                  & \qw & 
\qw                                 & \qw & \qw                                  & \qw &\ghost{QFT^\dagger}           & \qw & \qw \\
& \qw & \ghost{QFT}           & \qw & \gate{\theta_2^{a_{[0]}}} \qwx & \qw & \gate{\theta_1^{a_{[1]}}} \qwx & \qw & \gate{\theta_0^{a_{[2]}}}          & \qw & 
\qw                                 & \qw & \qw                                  & \qw &\ghost{QFT^\dagger}           & \qw & \qw \\
& \qw & \ghost{QFT}           & \qw & \gate{\theta_3^{a_{[0]}}} \qwx & \qw & \gate{\theta_2^{a_{[1]}}} \qwx & \qw & \gate{\theta_1^{a_{[2]}}} \qwx & \qw & \gate{\theta_0^{a_{[3]}}}         & \qw & \qw                                  & \qw &\ghost{QFT^\dagger}           & \qw & \qw \\
& \qw & \ghost{QFT}           & \qw & \gate{\theta_4^{a_{[0]}}} \qwx & \qw & \gate{\theta_3^{a_{[1]}}} \qwx & \qw & \gate{\theta_2^{a_{[2]}}} \qwx & \qw & \gate{\theta_1^{a_{[3]}}} \qwx & \qw & \gate{\theta_0^{a_{[4]}}} & \qw &\ghost{QFT^\dagger}           & \qw & \qw 
}
\end{eqnarray*}
\begin{eqnarray*}
\\
(c)
\\
\Qcircuit @C=.5em @R=.28em @!R {
& \qw & \gate{\theta_4}          &\qw&\qw\\
& \qw & \gate{\theta_3} \qwx &\qw&\qw\\
& \qw & \gate{\theta_2} \qwx &\qw&\qw\\ 
& \qw & \gate{\theta_1} \qwx &\qw&\qw\\
& \qw  & \ctrl{-1}                    &\qw&\qw
}
\raisebox{-3.5em}{\hspace{1mm}=\hspace{1mm}}
\Qcircuit @C=.5em @R=.43em {
& \qw & \gate{H} & \multigate{4}{GMS_{EXP}} & \multigate{3}{GMS^\dagger_{EXP}} & \gate{H} & \qw & \qw \\
& \qw & \gate{H} & \ghost{GMS_{EXP}}           & \ghost{GMS^\dagger_{EXP}}           & \gate{H} &\qw & \qw \\
& \qw & \gate{H} & \ghost{GMS_{EXP}}           & \ghost{GMS^\dagger_{EXP}}           & \gate{H} &\qw & \qw \\
& \qw & \gate{H} & \ghost{GMS_{EXP}}           & \ghost{GMS^\dagger_{EXP}}           & \gate{H} &\qw & \qw \\
& \qw & \gate{H} & \ghost{GMS_{EXP}}           & \qw                                                   & \gate{H} &\qw & \qw 
}
\end{eqnarray*}
\caption{\label{fig9}
GMS-based QFT and QFA circuits. (a) shows the QFT circuit, where $\theta_d$ denotes a phase rotation gate with the rotation angle $\pi/2^d$. (b) shows the QFA circuit, where $\theta_d^{a_{[j]}}$ denotes a phase rotation gate with the rotation angle $\pi/2^d$, where $a_{[j]}$ denotes the control qubit that corresponds to the $j^{\text{th}}$ bit value of the integer $a$ of the input state $|a\rangle$. (c) shows a subcircuit of the shape that repeatedly appears in (a) and (b), and how it may be implemented using only two GMS gates. The subscript $EXP$ of GMS$_{EXP}$ denotes the exponential drop off in the strength of the interaction, {\em i.e.}, $\chi_{ij} \sim \pi/2^{|i-j|}$. 
}
\end{figure}
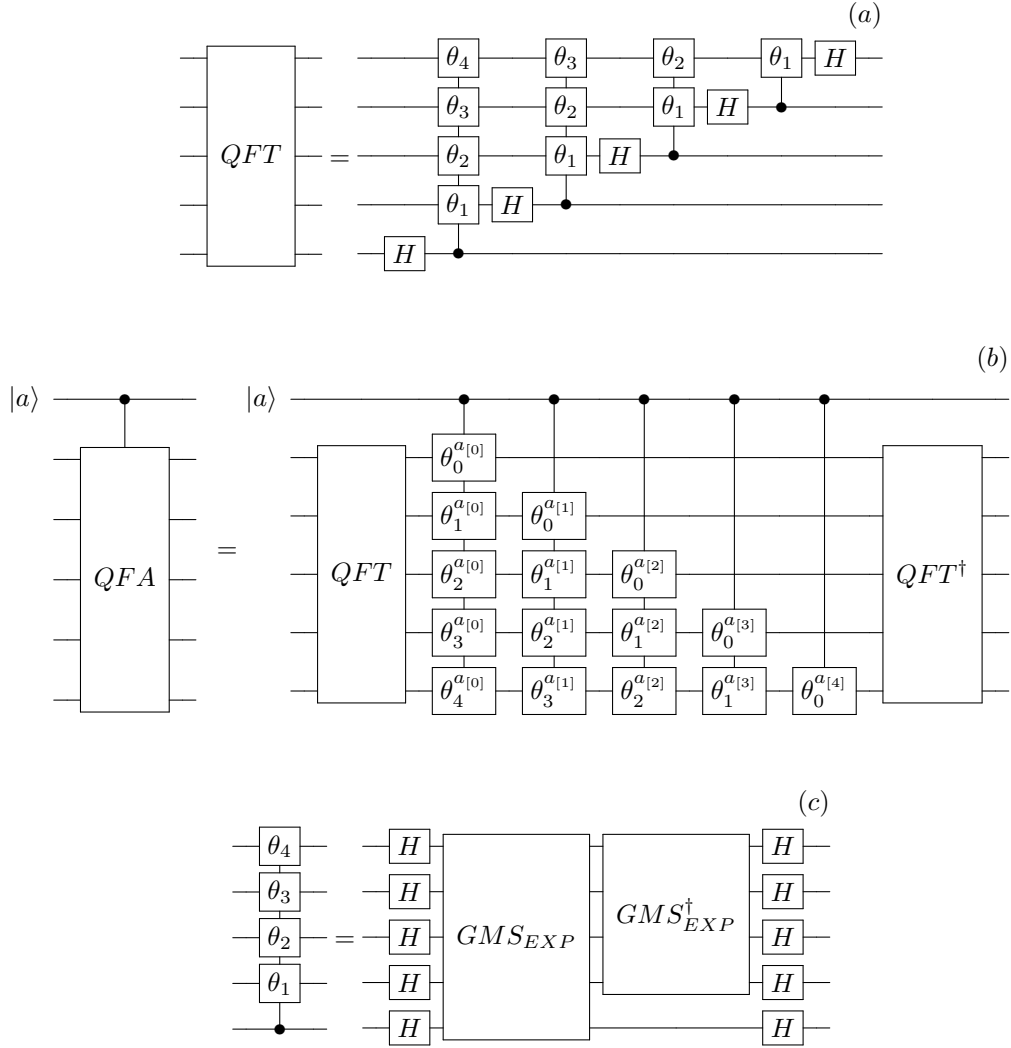

In case when the strength of the interaction falls off exponentially fast, as $\chi_{ij} \sim 2^{-|i-j|}$, it can be easily shown that the quantum Fourier transform (QFT) may be constructed efficiently using such global pulses.  Specifically, the efficient implementation uses just $2n$ global pulses, as opposed to $\frac{n(n-1)}{2}$ local two-qubit gates, for a QFT of size $n$.  This also enables an implementation of the quantum Fourier adder (QFA) \cite{Beau} with only a linear number of global gates, rather than superlinear, making the Fourier-based arithmetic circuits more competitive than the Boolean counterparts \cite{Cucc}.  Figs.~\ref{fig9} (a) and (b) show the QFT and QFA circuits.  Fig.~\ref{fig9}(c) illustrates how the GMS gates may be used to deliver the reduced gate count scaling in constructing the Fourier circuits.

Unfortunately, the exponential drop off in the strength of the interaction appears to be unnatural.  Instead, the decrease in the strength of the interaction as a power of the distance $d$ as $d^p$, where $p \in [0, 3]$ \cite{Monroe-PC, Hess} seems more realistic.  This leads to the question of how well the desired exponential drop off can be approximated with such physical-level global gates. \cite{NB-qip} provides an answer to this question.  Specifically, the quality of such Fourier circuits (QFT, QFA) is well preserved even when we alter the fundamental form of the signal strength.  For instance \cite{NB-qip}, when one replaces the exponential drop off, $\pi/2^d$, where $d$ is the distance between qubits, with a power law hierarchy, such as $\pi/d^{p}$, one may choose the power $p$ of the drop off power law such as to obtain the best possible quality of approximation.  In fact, it has been calculated numerically that the power $p_{\rm opt} = 1.4$ renders the maximum quality for the set of parameters considered in \cite{NB-qip}. Fortunately, $p=1.4$ is within the limits $p \in [0, 3]$ \cite{Monroe-PC, Hess}.

\begin{figure}[t]
\begin{tabular}{cc}
\includegraphics{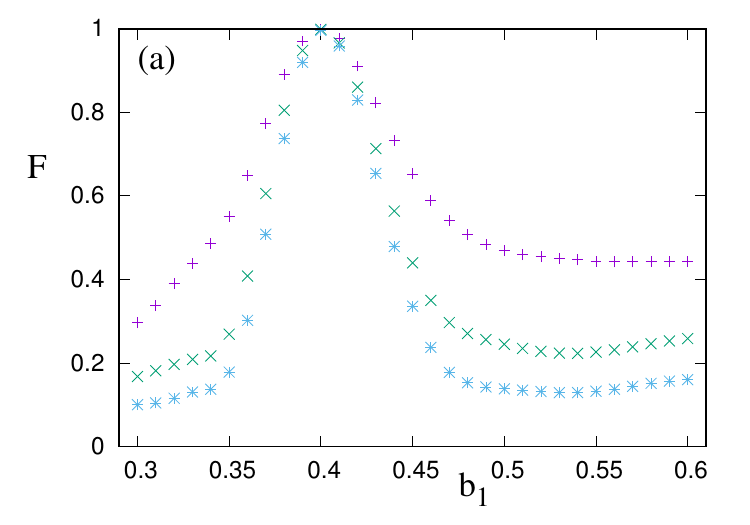} &
\includegraphics{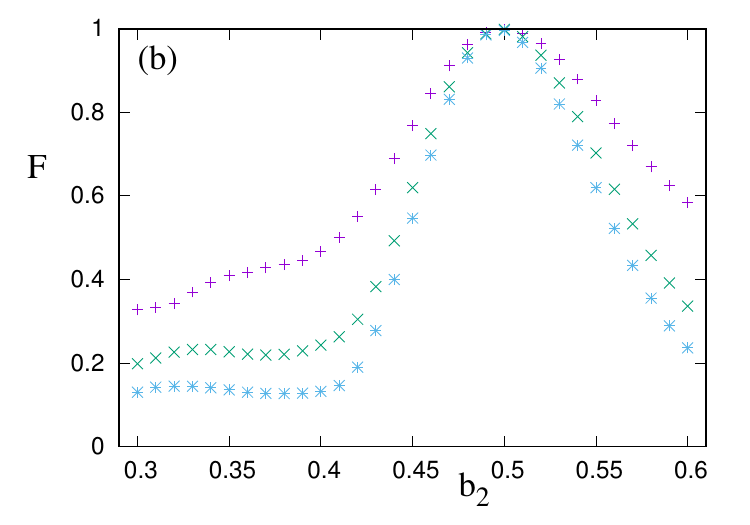} \\
\includegraphics{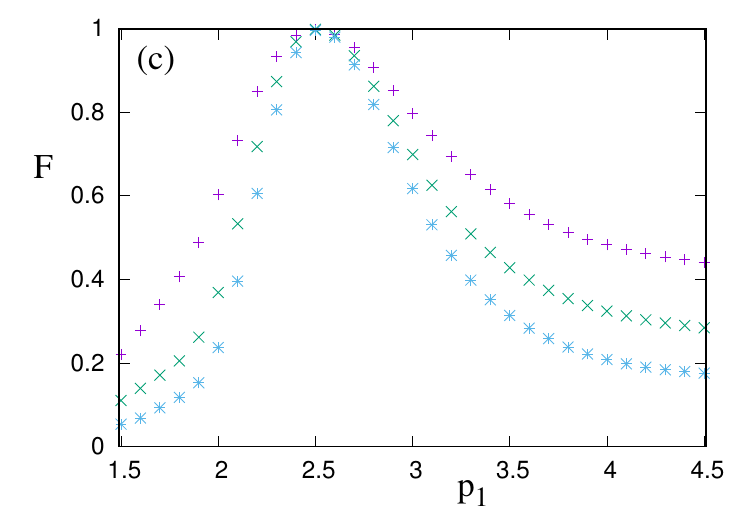} &
\includegraphics{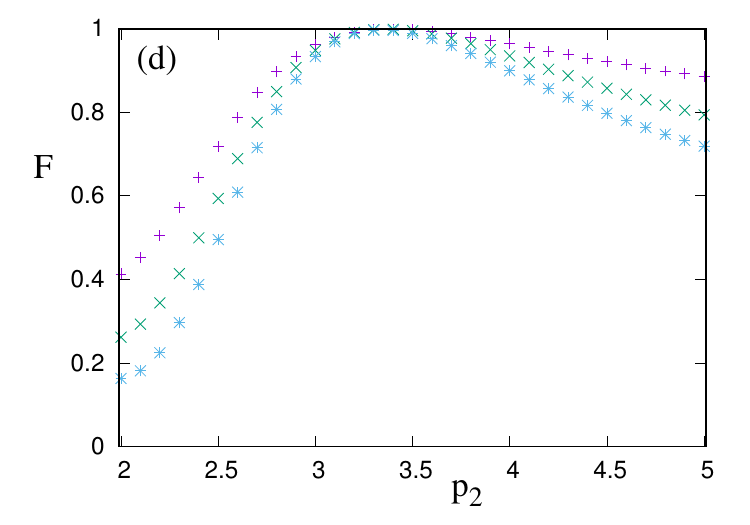} \\
\end{tabular}
\caption{\label{fig10}
Fidelity of power law QFT with $m=2$ near its analytically predicted optimum,
$b_1 = 0.4$, $b_2 = -0.5$, $p_1 = 2.5$, $p_2 = 3.4$.
In the order of (a)-(d), we fix all four parameters except for
(a) $b_1$, (b) $b_2$, (c) $p_1$, and (d) $p_2$.
$n=10$ (pluses), $n=12$ (crosses), and $n=14$ (asterisks).
}
\end{figure}

Motivated by the previous discussion, we next suggest an extended method of the power law approximation of the exponential drop off that is useful for quantum Fourier arithmetic circuits.  Specifically, we propose using a few GMS gates to approximate a single stage of the exponentially dropping interaction strength (see Fig.~\ref{fig9}(c)). This is in contrast to a simple replacement of the single-stage exponential drop off with a single-stage power law drop off, as was done in \cite{NB-qip}.  In particular, we numerically approximate the exponential drop off with a set of $m$ power law drop offs, as follows, 
\begin{equation}
\label{Pwr_Exp}
\frac{\pi}{2^j} \approx \sum_{i=1}^m \frac{\pi}{b_i j^{p_i}}.
\end{equation}
Since the circuit realization of each power law requires two GMS gates, the approximation by $m$ power laws amounts to a cost of $2m$ GMS gates.

Our goal is to numerically determine a set of $b_i$ and $p_i$ such that the term (\ref{Pwr_Exp}) minimizes the approximation error, in order to best match the exact exponential drop off, as seen in the quantum Fourier arithmetic circuits.  A straightforward generalization of the crude, yet efficient analytical works shown in \cite{NB-qip} reveals that the fidelity of the QFT may be approximated
by the term
\begin{equation}
\label{QFT_Fid_Pwr}
F_{QFT} \approx \exp\left\{ -\pi^2 \sum_{j=1}^{n} \frac{3(n-j)}{64} \left[ \frac{1}{2^j}
 - \left( \sum_{i=1}^m \frac{1}{b_i j^{p_i}} \right) \right]^2 \right\},
\end{equation}
meaning we obtain the best fidelity by minimizing the value of the sum in the exponent in (\ref{QFT_Fid_Pwr}). Minimizing the exponent in (\ref{QFT_Fid_Pwr}) analytically is a non-trivial task, and we thus resort to a numerical approximation. In particular, we restricted the search to $|b_i| \leq 0.6$ and $1.5 \leq p_i \leq 4$, closely following what may be achievable in the lab.  We find that for $m = 2$ the selection of the values $b_1 = 0.4$, $b_2 = -0.5$, $p_1 = 2.5$, $p_2 = 3.4$ results in the minimal exponent in (\ref{QFT_Fid_Pwr}), that is consistent with the peaks in fidelity observed in Fig.~\ref{fig8} for the sample cases of the QFT with $n = 10, 12,$ and $14$ qubits.  The peak fidelity $F_{peak} \approx 1$ demonstrates a high quality of the double-power approximation of the exponential drop off, making the efficient GMS-based construction an attractive alternative in experiments. 

We also conducted a similar numerical investigation for the QFA. This time, since $j=0$, $\pi$-rotation (see (\ref{Pwr_Exp})) in the QFA as shown in Fig.~\ref{fig9} needs to be implemented, we modify our power-law expansion according to
\begin{equation}
\label{Pwr_Exp2}
\frac{\pi}{2^j} \approx \sum_{i=1}^m \frac{\pi}{b_i (j+1)^{p_i}}.
\end{equation}
Once again, we numerically found the best choices of $(b,p)$ pairs, as in the previous case of the QFT with $m=2$, that result in the best performance.

\section{Summary of the results}\label{SecSum}

In Table \ref{tab:main} we summarized the advantage of using global entangling pulse enabled constructions developed in this work over best known circuitry relying on both local entangling control and, when available, global entangling control.  Columns \#q and \#eg show the number of physical qubits and the number of entangling gates needed to implement the operation specified by the column ``Operation'', with various approaches to the entangling control specified by the names of the multicolumns.  The benchmark functions used are Toffoli-$n$---the maximal size multiple control Toffoli gate over $n$ qubits, AQFT-$n$---approximate QFT over $n$ qubits, AQFA-$n$---approximate QFA of two $n$-bit numbers, and Tdistill---encoding circuit for the [[15,1,3]] code \cite{RHG07}.  We selected between 2-GMS gate and 4-GMS gate approximations of the circuit layers (see Fig. \ref{fig9}(c)) to best demonstrate the advantage over two-qubit local control, and obtained the two-qubit gate count for AQFT and AQFA circuits over local control such as to match the performance of GMS-enabled constructions.  We broke down the set of operations that benefit from GMS gates into three subsets---those suitable for near-term demonstration (selected by the virtue of relying on an already available number of qubits using a small number of entangling pulses; top circuit in the table), those targeted for next-generation machines (roughly, $10$ to $15$ qubits; second third of the table), and those applying to arbitrary $n$ (bottom third of the table).  Observe that for circuits suitable for the implementation over near-term and next-generation machines the advantage in the entangling gate count enabled by the global control is roughly by a factor $1.39$ to $3.67$, {\em i.e.}, it is substantial.  The minimal advantage shown is by a factor of $1.39$ for the circuit AQFA-$5$.  This circuit adds two 5-bit numbers and relies on the circuit layers with at most 5 two-qubit gates (see Fig. \ref{fig9}(b)).  Such layers are too short to show a significant advantage in approximating by 2 or 4 GMS gates, and the advantage becomes more pronounced as the number of qubits grows.  Specifically, the 20-qubit AQFA-10 already enjoys the optimization from 94 two-qubit local gates down to 53 entangling gates using a mix of global and local control, {\em i.e.}, by a factor of $1.77$.

\begin{table}[t] 
{\centering
\begin{tabular}{c|cc|cc|cc} 
Operation & \multicolumn{2}{|c|}{Local control} 		& \multicolumn{2}{c|}{Global control (best known)} & \multicolumn{2}{c}{Global and local control (ours)} \\
  			& \#q 	& \#eg 						& \#q 	& \#eg 				& \#q 	& \#eg \\ \hline
Toffoli-4 	& 5		& 11 \cite{FMLLDM17} 		& 4 	& 7 \cite{Ivanov}	& 5 & 3   \\ \hline
Toffoli-8   & 11 	& 35 \cite{FMLLDM17,M16} 	& 13 	& 33 \cite{MKH09,NC}& 11 & 15 \\
Toffoli-9   & 12	& 41 \cite{FMLLDM17,M16} 	& 15 	& 39 \cite{MKH09,NC}& 13 & 21 \\
Toffoli-10  & 14	& 47 \cite{FMLLDM17,M16} 	& 17 	& 45 \cite{MKH09,NC}& 14 & 21 \\
AQFT-10 	& 10 	& 30 						& N/A 	& N/A 				& 10 & 17 \\
AQFT-11 	& 11 	& 34 						& N/A 	& N/A 				& 11 & 19 \\
AQFT-12 	& 12 	& 38 						& N/A 	& N/A 				& 12 & 21 \\
AQFT-13 	& 13 	& 42 						& N/A 	& N/A 				& 13 & 23 \\
AQFT-14 	& 14 	& 46 						& N/A 	& N/A 				& 14 & 25 \\
AQFT-15 	& 15 	& 50 						& N/A 	& N/A 				& 15 & 27 \\
AQFA-5 		& 10	& 32 						& N/A 	& N/A 				& 10 & 23 \\
AQFA-6 		& 12	& 42 						& N/A 	& N/A 				& 12 & 29 \\
AQFA-7 		& 14	& 58 						& N/A 	& N/A 				& 14 & 35 \\
Tdistill 	& 15 	& 34 \cite{RHG07} 			& N/A 	& N/A 				& 15 & 10 \\ \hline
Toffoli-$n$ & $\lceil\frac{3n-3}{2}\rceil$ & $6n-13$ \cite{FMLLDM17,M16} & $2n-3$ & $6n-15$ \cite{MKH09,NC} & $\lceil\frac{3n-2}{2}\rceil$ & $6\lceil\frac{n}{2}\rceil-9$ \\
Stabilizer	& $n$ & $O\left(\frac{n^2}{\log(n)}\right)$ \cite{MR17,PMH08} & N/A & N/A & $n$ & $O(n)$ \\
\end{tabular} 
\caption{Advantages to the use of GMS gates.  The numbers for the global control enabled implementations of the Toffoli-$8..10$ were obtained by combining the 3-GMS implementation of the Toffoli-3 from \cite{MKH09} with the nested construction illustrated in \cite[Figure 4.10]{NC}.  The two-qubit gate counts for AQFT-$10..15$ and AQFA-$5..7$ circuits were obtained using known standard constructions, so as to match the approximation quality of our GMS-enabled implementations while using the minimal number of local gates.  A high number of N/A shown in the Table may suggest that not enough effort has been put into developing implementations over global control yet, highlighting one of the main messages of our paper.} \label{tab:main}
}
\end{table}

\section{Conclusion}\label{SecV}

In this paper, we studied the efficient use of a global entangling operator in realizing quantum circuitry of practical interest.  We developed a number of circuit equalities using GMS gates, improving the accessibility of global entangling gates in quantum circuit constructions.  Using various versions of the global entangling operator, we demonstrated the advantage in implementing stabilizer circuits, Toffoli-4 gate, Toffoli-$n$ gate, Quantum Fourier Transformation, and Quantum Fourier Adder circuits.  In each of the above, our circuits outperform best known circuitry in the scenario when the control is given by the two-qubit local addressable gates.  Our conclusion is as follows: we believe that the control by a global entangling gate (an analogue of single instruction, multiple data classical architecture) could be a helpful complement to the control by addressable two-qubit local gates.

\section*{Acknowledgements}

Authors thank Prof. Kenneth Brown from Georgia Institute of Technology and Prof. Christopher Monroe from the University of Maryland -- College Park for their discussions and help in the preparation of this manuscript.

This material was partially based on work supported by the National Science Foundation during DM's assignment at the Foundation. Any opinion, finding, and conclusions or recommendations expressed in this material are those of the authors and do not necessarily reflect the views of the National Science Foundation.

YN acknowledges support from ARO MURI award W911NF-16-1-0349.

\end{document}